\documentclass[runningheads,a4paper]{llncs}

\usepackage{amssymb}
\usepackage{amsmath}
\usepackage{graphicx}
\usepackage[table,pdftex]{xcolor}
\usepackage{xspace}
\usepackage[colorlinks=true,citecolor=blue,urlcolor=black]{hyperref}
\usepackage{lineno}
\usepackage{subfig}
\usepackage[colorinlistoftodos]{todonotes}
\usepackage{wrapfig}
\usepackage[misc]{ifsym}
\usepackage{amsfonts}
\usepackage{booktabs}
\usepackage{multirow}
\usepackage{siunitx}

\begin{document}

\mainmatter  % start of an individual contribution
\title{VS-Net: Variable splitting network for accelerated parallel MRI reconstruction }
\titlerunning{VS-Net: Variable splitting deep neural network}

% remove asterisk
\makeatletter
\def\thanks#1{\protected@xdef\@thanks{\@thanks
        \protect\footnotetext{#1}}}
\makeatother

\mainmatter  % start of an individual contribution
\title{VS-Net: Variable splitting network for accelerated parallel MRI reconstruction }
\titlerunning{VS-Net: Variable splitting deep neural network}

\newcommand{\corrauth}{\textsuperscript{(\Letter)}}
\author{Jinming Duan$^{\dagger1,2}$\corrauth\thanks{$^\dagger$contributed equally. }
, Jo Schlemper$^{\dagger2}$, Chen Qin$^{2}$, Cheng Ouyang$^{2}$, \\Wenjia Bai$^{2}$, Carlo Biffi$^{2}$, Ghalib Bello$^{3}$, Ben Statton$^{3}$, \\Declan P O'Regan$^{3}$, Daniel Rueckert$^{2}$}
\authorrunning{J Duan, et al.}
\institute{$^1$School of Computer Science, University of Birmingham, Birmingham, UK\\
\email{j.duan@cs.bham.ac.uk}\\
$^2$Biomedical Image Analysis Group, Imperial College London, London, UK\\
$^3$MRC London Institute of Medical Sciences, Imperial College London, London, UK\\}

\maketitle

%\linenumbers %remove before submission

% !TeX root=../paper.tex

\begin{abstract}
In this work, we propose a deep learning approach for parallel magnetic resonance imaging (MRI) reconstruction, termed a variable splitting network (VS-Net), for an efficient, high-quality reconstruction of undersampled multi-coil MR data. We formulate the generalized parallel compressed sensing reconstruction as an energy minimization problem, for which a variable splitting optimization method is derived. Based on this formulation we propose a novel, end-to-end trainable deep neural network architecture by unrolling the resulting iterative process of such variable splitting scheme. VS-Net is evaluated on complex valued multi-coil knee images for 4-fold and 6-fold acceleration factors. We show that VS-Net outperforms state-of-the-art deep learning reconstruction algorithms, in terms of reconstruction accuracy and perceptual quality. Our code is publicly available at \href{https://github.com/j-duan/VS-Net}{https://github.com/j-duan/VS-Net}.

\end{abstract}

% !TeX root=../paper.tex

\section{Introduction}

Magnetic resonance imaging (MRI) is an important diagnostic and research tool in many clinical scenarios. However, its inherently slow data acquisition process is often problematic. To accelerate the scanning process, parallel MRI (p-MRI) and compressed sensing MRI (CS-MRI) are often employed. These methods are designed to facilitate fast reconstruction of high-quality, artifact-free images from minimal $k$-space data. Recently, deep learning approaches for MRI reconstruction \cite{schlemper2018deep,sun2016deep,hammernik2018learning,aggarwal2019modl,han2018k,akccakaya2018scan,jin2019self,mardani2019deep,tezcan2018mr,zhang2018multi} have demonstrated great promises for further acceleration of MRI acquisition. However, not all of these techniques \cite{schlemper2018deep,sun2016deep,han2018k,jin2019self,mardani2019deep} are able to exploit parallel image acquisitions which are common in clinical practice.

In this paper, we investigate accelerated p-MRI reconstruction using deep learning. We propose a novel, end-to-end trainable approach for this task which we refer to as a variable splitting network (VS-Net). VS-Net builds on a general parallel CS-MRI concept, which is formulated as a multi-variable energy minimization process in a deep learning framework. It has three computational blocks: a denoiser block, a data consistency block and a weighted average block. The first one is a denoising convolutional neural network (CNN), while the latter two have point-wise closed-form solutions. As such, VS-Net is computationally efficient yet simple to implement. VS-Net accepts complex-valued multi-channel MRI data and learns all parameters automatically during offline training. In a series of experiments, we monitor reconstruction accuracies obtained from varying the number of stages in VS-Net. We studied different parameterizations for weight parameters in VS-Net and analyzed their numerical behaviour. We also evaluated VS-Net performance on a multi-coil knee image dataset for different acceleration factors using the Cartesian undersampling patterns, and showed improved image quality and preservation of tissue textures.

To this end, we point out the differences between our method and related works in \cite{schlemper2018deep,sun2016deep,hammernik2018learning,aggarwal2019modl,mardani2019deep}, as well as highlight our novel contributions to this area. First, data consistency (DC) layer introduced in \cite{schlemper2018deep} was designed for single-coil images. MoDL \cite{aggarwal2019modl} extended the cascade idea of \cite{schlemper2018deep} to a multi-coil setting. However, its DC layer implementation was through iteratively solving a linear system using the conjugate gradient in their network, which can be very complicated. In contrast, DC layer in VS-Net naturally applies to multi-coil images, and is also a point-wise, analytical solution, making VS-Net both computationally efficient and numerically accurate. Variational network (VN) \cite{hammernik2018learning} and \cite{mardani2019deep} were applicable to multi-coil images. However, they were based gradient-descent optimization and proximal methods respectively, which does not impose the exact DC. Compared ADMM-net \cite{sun2016deep} to VS-Net, the former was also only applied to single-coil images. Moreover, ADMM-Net was derived from the augmented Lagrangian method (ALM), while VS-Net uses a penalty function method, which results in a simpler network architecture. ALM introduces Lagrange multipliers to weaken the dependence on penalty weight selection. While these weights can be learned automatically in network training, the need for a network with a more complicated ALM is not clear. In ADMM-Net and VN, the regularization term was defined via a set of explicit learnable linear filter kernels. In contrast, VS-Net treats regularization implicitly in a CNN-denoising process. Consequently, VS-Net has the flexibility of using varying advanced denoising CNN architectures while avoiding expensive dense matrix inversion - a encountered problem in ADMM-Net. A final distinction is that the effect of different weight parameterizations is studied in VS-Net, while this was not investigated in the aforementioned works.

\section{VS-Net for accelerated p-MRI reconstruction}
\label{sec:method}
\textbf{General CS p-MRI model:} Let $m \in {\mathbb{C}^N}$ be a complex-valued MR image stacked as a column vector and $y_i \in {\mathbb{C}^M}$ ($M < N$) be the under-sampled $k$-space data measured from the $i$th MR receiver coil. Recovering $m$ from $y_i$ is an ill-posed inverse problem. According to compressed sensing (CS) theory, one can estimate the reconstructed image $m$ by minimizing the following unconstrained optimisation problem:   
\begin{equation} \label{eq:sense}
\mathop {\min }\limits_m \left\{ {\frac{\lambda }{2}\sum\limits_{i = 1}^{{n_c}} {\| {{\cal D} {\cal F}{S_i}m - {y_i}} \|_2^2}  + {\cal R}\left( m \right)} \right\},
\end{equation}
In the  data fidelity term (first term), $n_c$ denotes the number of receiver coils, ${\cal D} \in {\mathbb{R}^{M \times N}}$ is the sampling matrix that zeros out entries that have not been acquired, ${\cal F } \in {\mathbb{C}^{N \times N}}$ is the Fourier transform matrix, ${S_i } \in {\mathbb{C}^{N \times N}}$ is the $i$th coil sensitivity, and $\lambda$ is a model weight that balances the two terms. Note that the coil sensitivity $S_i$ is a diagonal matrix, which can be pre-computed from the fully sampled $k$-space center using the E-SPIRiT algorithm \cite{uecker2014espirit}. The second term is a general sparse regularization term, e.g. (nonlocal) total variation \cite{lu2016implementation,lu2019graph}, total generalized variation \cite{lu2016implementation,duan2016denoising} or the $\ell 1$ penalty on the discrete wavelet transform of $m$ \cite{liu2019undersampled}. \\

\noindent \textbf{Variable splitting:} In order to optimize (\ref{eq:sense}) efficiently, we develop a variable splitting method. Specifically, we introduce the auxiliary splitting variables $u  \in {\mathbb{C}^N} $ and $\{x_i  \in {\mathbb{C}^N}\}_{i=1}^{n_c}$, converting (\ref{eq:sense}) into the following equivalent form
\begin{equation} \nonumber 
\mathop {\min }\limits_{m,u,x_i}  {\frac{\lambda }{2}\sum\limits_{i = 1}^{{n_c}} {\| {{\cal D} {\cal F}x_i - {y_i}} \|_2^2}  + {\cal R}\left( u \right)} \; s.t.\;m=u, \;S_im=x_i, \;\forall i \in \left\{ {1,2,...,{n_c}} \right\}.
\end{equation} 
The introduction of the first constraint $m=u$ decouples $m$ in the regularization from that in the data fidelity term so that a denoising problem can be explicitly formulated (see Eq. (\ref{eq:subproblems}) top). The introduction of the second constraint $S_im=x_i$ is also crucial as it allows decomposition of $S_im$ from ${\cal D} {\cal F}{S_i}m $ in the data fidelity term such that no dense matrix inversion is involved in subsequent calculations (see Eq. (\ref{eq:solutions}) middle and bottom). Using the penalty function method, we then add these constraints back into the model and minimize the single problem
\begin{equation}\label{eq:splitsense}
\mathop {\min }\limits_{m,u,x_i}  {\frac{\lambda }{2}\sum\limits_{i = 1}^{{n_c}} {\| {{\cal D} {\cal F}x_i - {y_i}} \|_2^2}  + {\cal R}\left( u \right)}  + \frac{\alpha }{2}\sum\limits_{i = 1}^{{n_c}} {\| {{x_i} - {S_i}m} \|_2^2}  + \frac{\beta }{2}\| {u - m} \|_2^2\ , 
\end{equation}
where $\alpha$ and $\beta$ are introduced penalty weights. To minimize (\ref{eq:splitsense}), which is a multi-variable optimization problem, one needs to alternatively optimize $m$, $u$ and $x_i$ by solving the following three subproblems:
\begin{equation} \label{eq:subproblems}
\left\{ \begin{array}{l}
{u^{k + 1}} = \mathop {\arg \min }\limits_u \frac{\beta }{2}\| {u - {m^k}} \|_2^2 + {\cal R}\left( u \right)\\
x_i^{k + 1} = \mathop {\arg \min }\limits_{{x_i}} \lambda \sum\nolimits_{i = 1}^{{n_c}} {\| {{\cal D} {\cal F}{x_i} - {y_i}} \|_2^2}  + \frac{\alpha }{2}\sum\nolimits_{i = 1}^{{n_c}} {\| {{x_i} - {S_i}{m^k}} \|_2^2} \\
m^{k+1} = \mathop {\arg \min }\limits_m \frac{\alpha }{2}\sum\nolimits_{i = 1}^{{n_c}} {\| {x_i^{k + 1} - {S_i}m} \|_2^2}  + \frac{\beta }{2}\| {{u^{k + 1}} - m} \|_2^2
\end{array} \right.,
\end{equation}
Here $k \in \{1,2,...,n_{it}\}$ denotes the $k$th iteration. An optimal solution ($m^*$) may be found by iterating over ${u^{k + 1}}$, $x_i^{k+1}$ and ${m^{k + 1}}$ until convergence is achieved or the number of iterations reaches $n_{it}$. An initial solution to these subproblems can be derived as follows
\begin{equation} \label{eq:solutions}
\left\{ \begin{array}{l}
{u^{k + 1}} = denoiser({m^k})\\
x_i^{k + 1} = {{\cal F}^{ - 1}}( {{{( {\lambda {{\cal D} ^T}{\cal D}  + \alpha I } )}^{-1}}( {\alpha {\cal F}{S_i}{m^k} + \lambda{{\cal D} ^T}{y_i}} )})\;\;\; \forall i \in \{1,2,...,{n_c}\}\\
{m^{k + 1}} = {( {\beta I + \alpha\sum\nolimits_{i = 1}^{{n_c}} {S_i^H{S_i}} } )^{-1}}({\beta {u^{k + 1}} + \alpha \sum\nolimits_{i = 1}^{{n_c}} {S_i^H} x_i^{k + 1}})
\end{array} \right..
\end{equation}
Here $S_i^H$ is the conjugate transpose of $S_i$ and $I$ is the identity matrix of size $N$ by $N$. ${{\cal D} ^T}{\cal D}$ is a diagonal matrix of size $N$ by $N$, whose diagonal entries are either zero or one corresponding to a binary sampling mask. ${\cal D}^Ty_i$ is an $N$-length vector, representing the $k$-space measurements ($i$th coil) with the unsampled positions filled with zeros. In this step we have turned the original problem (\ref{eq:sense}) into a denoising problem (denoted by $denoiser$) and two other equations, both of which have closed-form solutions that can be computed point-wise due to the nature of diagonal matrix inversion. We also note that the middle equation efficiently imposes the consistency between $k$-space data and image space data coil-wisely, and the bottom equation simply computes a weighted average of the results obtained from the first two equations. Next, we will show an appropriate network architecture can be derived by unfolding the iterations of (\ref{eq:solutions}). 
\begin{figure}[h!] 
\vspace{-10pt}
\centering  
{\includegraphics[width=0.99\textwidth]{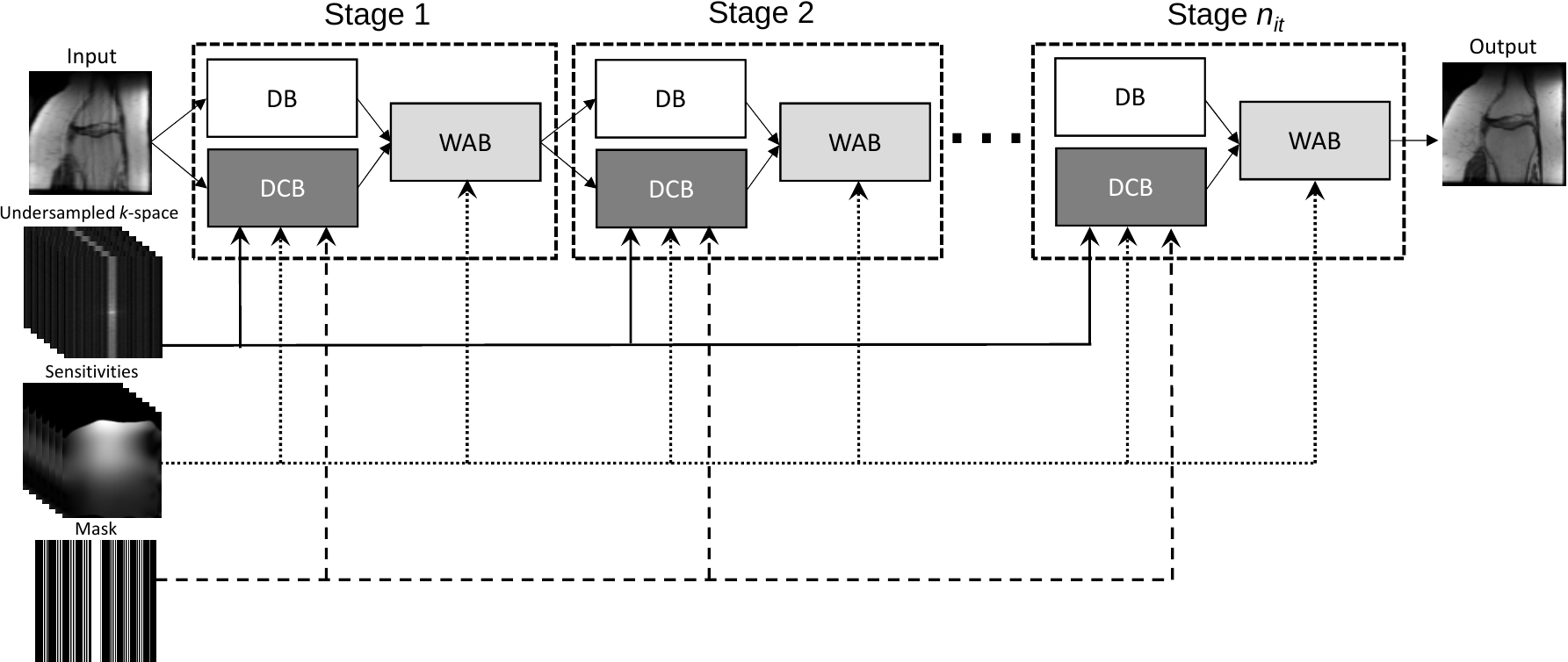}}\\
\vspace{-5pt}
\caption{Overall architecture of the proposed variable splitting network (VS-Net). }
\label{fig:network}
\vspace{-10pt}
\end{figure}

\begin{figure}[h!] 
\vspace{-10pt}
\centering  
{\includegraphics[width=0.99\textwidth]{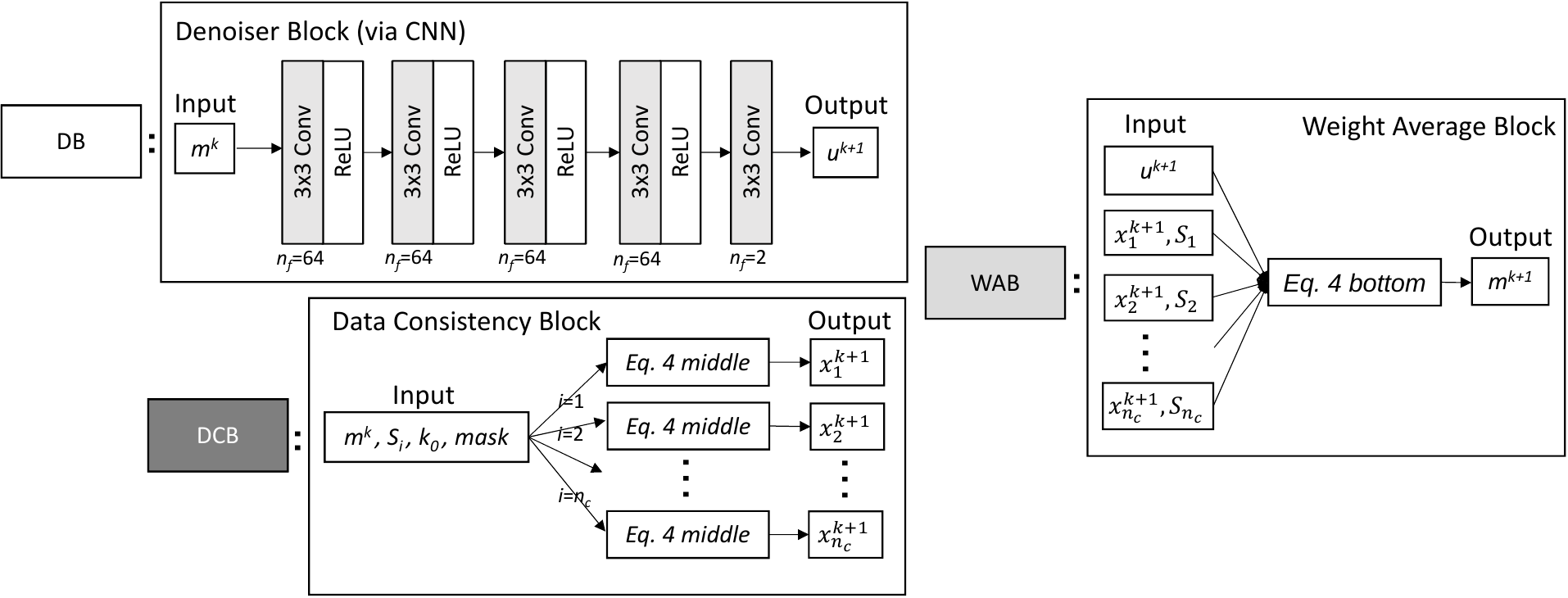}}\\
\vspace{-5pt}
\caption{Detailed structure of each block in VS-net. DB, DCB and WAB stand for Denoiser Block, Data Consistency Block and Weighted Average Block, respectively.}
\label{fig:blocks}
\vspace{-10pt}
\end{figure}

\noindent \textbf{Network architecture:} We propose a deep cascade network that naturally integrates the iterative procedures in Eq. (\ref{eq:solutions}). Fig.~\ref{fig:network} depicts the network architecture. Specifically, one iteration of an iterative reconstruction is related to one stage in the network. In each stage, there are three blocks: denoiser block (DB), data consistency block (DCB) and weighted average block (WAB), which respectively correspond to the three equations in (\ref{eq:solutions}). The network takes four inputs: 1) the single sensitivity-weighted undersampled image which is computed using $\sum_i^{n_c} S_i^H {\cal F}^{-1}{\cal D}^T y_i$; 2) the pre-computed coil sensitivity maps $\{S_i\}_{i=1}^{n_c}$; 3) the binary sampling mask ${{\cal D} ^T}{\cal D}$; 4)  the undersampled $k$-space data $\{{\cal D}^Ty_i\}_{i=1}^{n_c}$. Note that the sensitivity-weighted undersampled image is only used once for DB and DCB in Stage 1. In contrast, $\{{\cal D}^Ty_i\}_{i=1}^{n_c}$, $\{S_i\}_{i=1}^{n_c}$ and the mask are required for WAB and DCB at each stage (see Fig.~\ref{fig:network} and \ref{fig:blocks}). As the network is guided by the iterative process resulting from the variable splitting method, we refer to it as a Variable Splitting Network (VS-Net).

%are retrospectively computed from the fully sampled raw data $\{f_i \in {\mathbb{C}^N}\}_{i=1}^{n_c} $, the pre-computed coil sensitivity maps $\{S_i\}_{i=1}^{n_c}$, the binary sampling mask, as well as the undersampled $k$-space data $\{{\cal D}^Ty_i\}_{i=1}^{n_c}$. 
%
%First, the single sensitivity-weighted undersampled image is fed to DB and DCB in Stage 1. $\{{\cal D}^Ty_i\}_{i=1}^{n_c}$, $\{S_i\}_{i=1}^{n_c}$ and the mask are required for WAB and DCB at each stage (see Fig.~\ref{fig:network} and \ref{fig:blocks}). As the network is guided by the iterative process resulting from the variable splitting method, we refer to it as a variable splitting network (VS-Net).

In Fig.~\ref{fig:blocks}, we illustrate the detailed structures of key building blocks of the network (DB, DCB and WAB) at Stage $k$ in VS-Net. In DB, we intend to denoise a complex-valued image with a convolutional neural network (CNN). To handle complex values, we stack real and imaginary parts of the undersampled input into a real-valued two-channel image. ReLU's are used to add nonlinearities to the \textit{denoiser} to increase its denoising capability. Note that while we use a simple CNN here, our setup allows for incorporation of more advanced denoising CNN architectures. In DCB, $m^k$ from the upstream block, $\{S_i\}_{i=1}^{n_c}$, $\{k_i\}_{i=1}^{n_c}$ (i.e. the undersampled $k$-space data of all coils) and the mask are taken as inputs, passing through the middle equation of (\ref{eq:solutions}) and outputting $\{x_i^{k+1}\}_{i=1}^{n_c}$. The outputs $u^{k+1}$ and $\{x_i^{k+1}\}_{i=1}^{n_c}$ from DCB and WAB, concurrently with the coil sensitivity maps, are fed to WAB producing $m^{k+1}$, which is then used as the input to DB and DCB in the next stage in VS-Net. Due to the existence of analytical solutions, no iteration is required in WAB and DCB. Further, all the computational operations in WAB and DCB are point-wise. These features make the calculations in the two blocks simple and efficient. The process proceeds in the same manner until Stage $n_{it}$ is reached. \\

\noindent \textbf{Network loss and parameterizations:} Training the proposed VS-Net is another optimization process, for which a loss function must be explicitly formulated. In MR reconstruction, the loss function often defines the similarity between the reconstructed image and a clean, artifact-free reference image. For example, a common choice for the loss function used in this work is the mean squared error (MSE), given by 
\begin{equation} \label{eq:MSE}
{\cal L}({\bf{\Theta}}) = \min_{\bf{\Theta}} \frac{1}{{2{n_i}}}\sum\limits_{i = 1}^{{n_i}} \| {{m_i^{{n_{it}}}}( {\bf{\Theta}} ) - {g_i}} \|_2^2,
\end{equation}
where $n_i$ is the number of training images, and $g$ is the reference image, which is a sensitivity-weighted fully sampled image computed by $\sum_j^{n_c} S_j^H {\cal F}^{-1}f_j$. Here $f_j$ represents the fully sampled raw data of the $j$th coil. ${\bf{\Theta}}$ above are the network parameters ${\bf{\Theta}}$ to be learned. In this work we study two parameterizations for ${\bf{\Theta }}$, i.e., ${\bf{\Theta^1}} = \left\{ {\{{{\bf{{W}}}^l}\}_{l = 1}^{{n_{it}}}, \lambda, \alpha, \beta} \right\}$ and ${\bf{\Theta^2}} = \left\{{\{ {{\bf{{W}}}^l, \lambda^l, \alpha^l, \beta^l}\}_{l = 1}^{{n_{it}}}} \right\}$. Here $\{ {{\bf{{W}}}^l}\}_{l = 1}^{{n_{it}}}$ are learnable parameters for all ($n_{it}$) CNN denoising blocks in VS-Net. Moreover, in both cases we also make the data fidelity weight $\lambda$ and the penalty weights $\alpha$ and $\beta$ learnable parameters. In contrast, for ${\bf{\Theta^1}}$ we let the weights $\lambda$, $\alpha$ and $\beta$ be shared by the WABs and DCBs across VS-Net, while for ${\bf{\Theta^2}}$ each WAB and DCB have their own learnable weights. Since all the blocks are differentiable, backpropagation (BP) can be effectively employed to minimize the loss with respect to the network parameters ${\bf{\Theta}}$ in an end-to-end fashion. 

\section{Experiments results}
\noindent \textbf{Datasets and training details:} We used a publicly available clinical knee dataset\footnote{\href{https://github.com/VLOGroup/mri-variationalnetwork}{https://github.com/VLOGroup/mri-variationalnetwork}} in \cite{hammernik2018learning}, which has the following 5 image acquisition protocols: coronal proton-density (PD), coronal fat-saturated PD, axial fat-saturated T$_2$, sagittal fat-saturated T$_2$ and sagittal PD. For each acquisition protocol, the same 20 subjects were scanned using an off-the-shelf 15-element knee coil. The scan of each subject cover approximately 40 slices and each slice has 15 channels ($n_c=15$). Coil sensitivity maps provided in the dataset were precomputed from a data block of size 24 $\times$ 24 at the center of fully sampled $k$-space using BART \cite{uecker2013software}. 
For training, we retrospectively undersampled the original $k$-space data for 4-fold and 6-fold acceleration factors (AF) with Cartesian undersampling, sampling 24 lines at the central region. For each acquisition protocol, we split the subjects into training and testing subsets (each with sample size of 10), and trained VS-Net to reconstruct each slice in a 2D fashion. The network parameters was optimized for 200 epochs, using Adam with learning rate $10^{-3}$ and batch size 1. We used PSNR and SSIM as quantitative performance metrics.

\begin{figure}[h!]
\vspace{-10pt}
\centering  
{\includegraphics[width=1\textwidth]{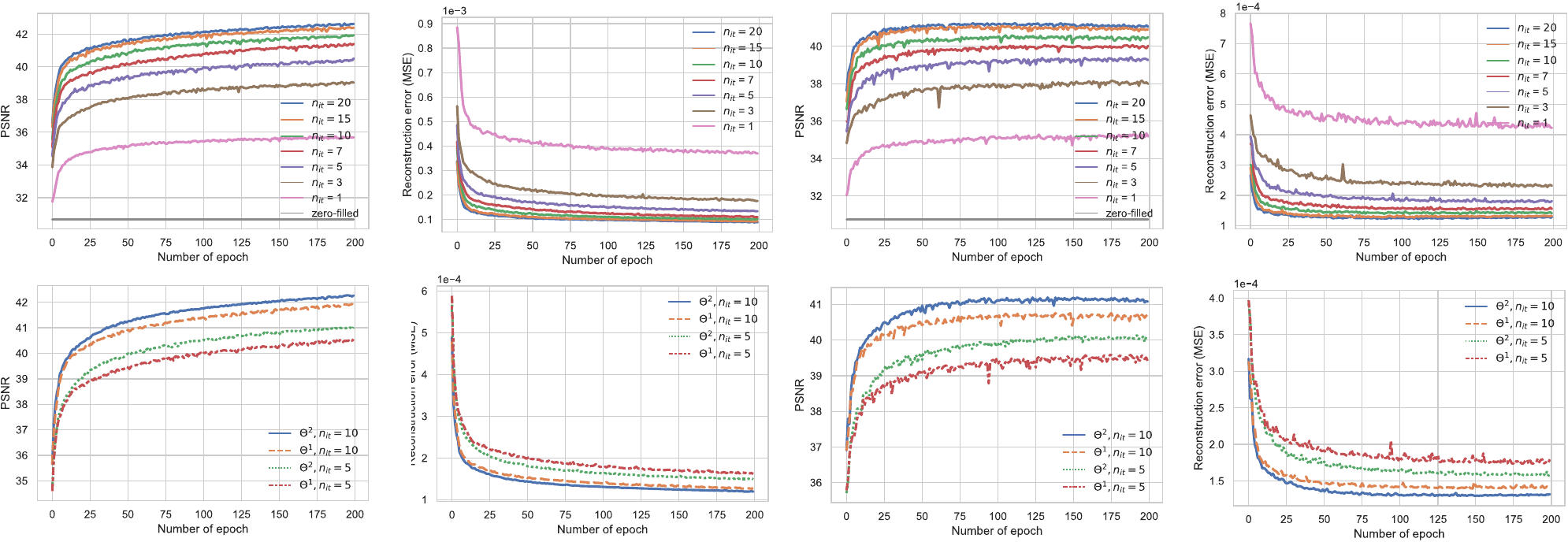}}\\
\vspace{-5pt}
\caption{Quantitative measures versus number of epochs at training (first two columns) and testing (last two columns). 1st row shows the network performance using different stage numbers. 2nd column shows the network performance using different parameterizations of ${\bf{\Theta}}$ in the loss (\ref{eq:MSE}).}
\label{fig:curves}
\vspace{-20pt}
\end{figure}

\textbf{Parameter behaviour:} To show the impact of the stage number $n_{it}$ (see Fig~\ref{fig:network}), we first experiment on the subjects under the coronal PD protocol with 4-fold AF. We set $n_{it} = \{1,3,5,7,10,15,20\}$ and plotted the training and testing quantitative curves versus the number of epochs in the upper portion of Fig~\ref{fig:curves}. As the plots show, increasing the stage number improves network performance. This is obvious for two reasons: 1) as number of parameters increases, so does the network's learning capability; 2) the embedded variable splitting minimization is an iterative process, for which sufficient iterations (stages) are required to converge to an ideal solution. We also found that: i) the performance difference between $n_{it} = 15$ and $n_{it} = 20$ is negligible as the network gradually converges after $n_{it} = 15$; ii) there is no overfitting during network training despite the use of a relatively small training set. Second, we examine the network performance when using two different parameterizations: ${\bf{\Theta^1}} = \left\{ {\{ {{\bf{{W}}}^l}\}_{l = 1}^{{n_{it}}},\lambda, \alpha, \beta } \right\}$ and ${\bf{\Theta^2}} = \left\{ {\{ {{\bf{{W}}}^l, \lambda^l, \alpha^l, \beta^l}\}_{l = 1}^{{n_{it}}} } \right\}$. For a fair comparison, we used the same initialization for both parameterizations and experimented with two cases $n_{it} = \{5,10\}$. As shown in the bottom portion of Fig~\ref{fig:curves}, in both cases the network with ${\bf{\Theta^1}}$ performs slightly worse than the one with ${\bf{\Theta^2}}$. In penalty function methods, a penalty weight is usually shared (fixed) across iterations. However, our experiments indicated improved performance if the model weights ($\lambda$, $\alpha$ and $\beta$) are non-shared or adaptive at each stage in the network.  

\begin{figure}[h!] 
\vspace{-10pt}
\centering  
{\includegraphics[width=1\textwidth]{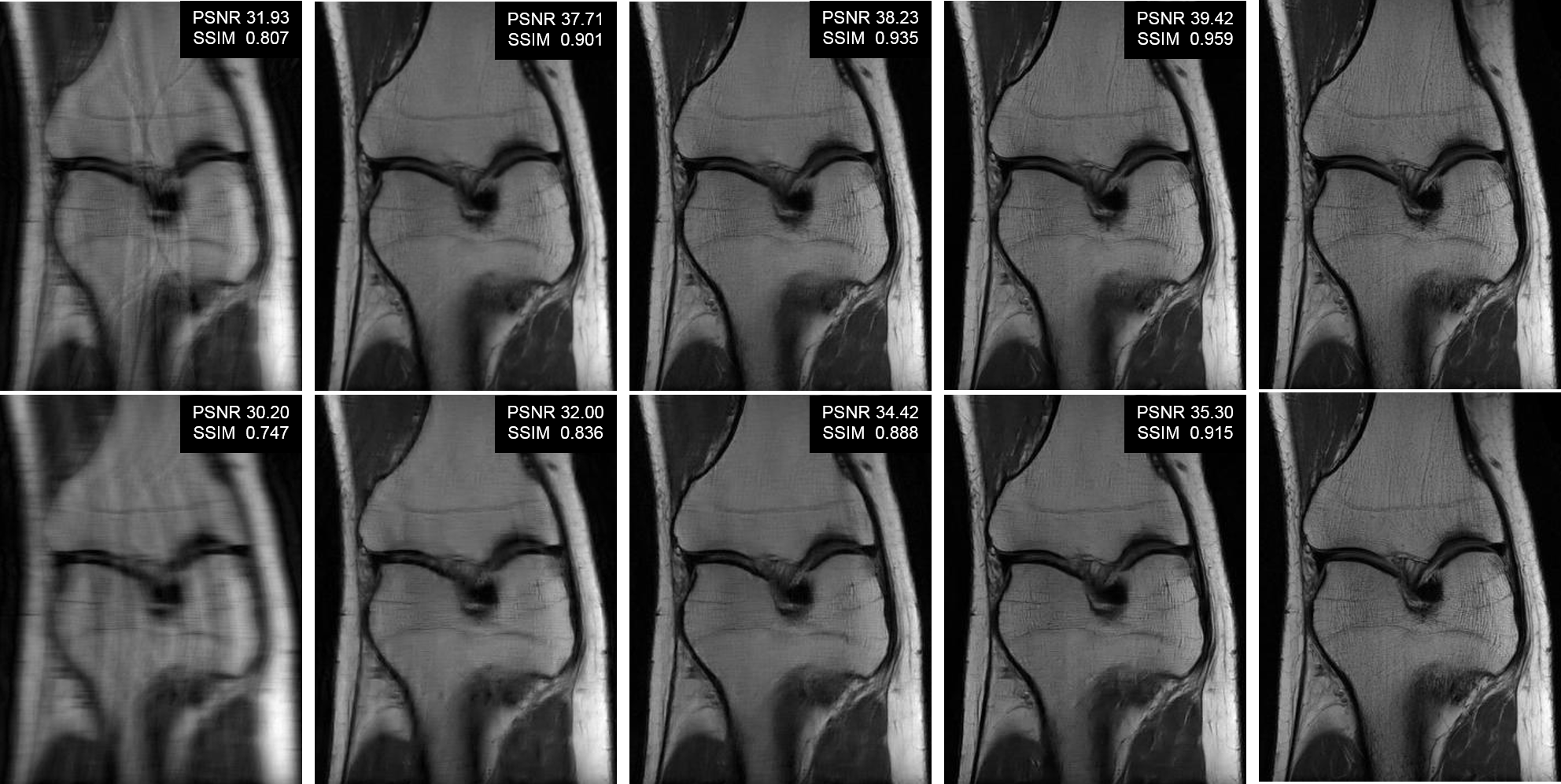}}\\
\vspace{-5pt}
\caption{Visual comparison of results obtained by different methods for Cartesian undersampling with AF 4 (top) and 6 (bottom). From left to right: zero-filling results, $\ell1$-SPIRiT results, VN results, VS-Net results, and ground truth. Click \href{http://www.cs.bham.ac.uk/~duanj/moive/more_visual_comparison.pdf}{here} for more visual comparison.}
\label{fig:visual}
\vspace{-17pt}
\end{figure}

\textbf{Numerical comparison:} We compared our VS-Net with the iterative $\ell1$-SPIRiT \cite{murphy2012fast} and the variational network (VN) \cite{hammernik2018learning}, with the zero-filling reconstruction as a reference. For VS-Net, we used $n_{it}$ (10) and $\bf{\Theta}^2$, although the network's performance can be further boosted with a larger $n_{it}$. For VN, we carried out training using mostly default hyper-parameters from \cite{hammernik2018learning}, except for the batch size, which (using original image size) was set to 1 to better fit GPU memory. For both VS-Net and VN, we trained a separate model for each protocol, resulting in a total of 20 models for the 4-fold and 6-fold AFs. In Table~\ref{tb:compare}, we summarize the quantitative results obtained by these methods. As is evident, learning-based methods VS-Net and VN outperformed the iterative $\ell1$-SPIRiT. VN produced comparable SSIMs to VS-Net in some scenarios. The resulting PNSRs were however lower than that of VS-Net for all acquisition protocols and AFs, indicating the superior numerical performance of VS-Net. In Fig~\ref{fig:visual}, we present a visual comparison on a coronal PD image reconstruction for both AFs. Apart from zero-filling, all methods removed aliasing artifacts successfully. Among $\ell1$-SPIRiT, VN and VS-Net, VS-Net recovered more small, textural details and thus achieved the best visual results, relative to the ground truth. The quantitative metrics in Fig~\ref{fig:visual} further show that VS-Net is the best.
\begin{table}[h!] 
\vspace{-20pt}
\centering
\caption{Quantitative results obtained by different methods on the test set including $\sim$2000 image slices across 5 acquisition protocols. Each metric was calculated on $\sim$400 image slices, and mean $\pm$ standard deviation are reported.}
\resizebox{\textwidth}{!}{
\begin{tabular}{|c|c|c|c|c|c|}
\hline
                                      &              & \multicolumn{2}{c|}{4-fold AF} & \multicolumn{2}{c|}{6-fold AF} \\ \hline
Protocol                              & Method       & PSNR           & SSIM          & PSNR           & SSIM          \\ \hline
\multirow{4}{*}{Coronal fat-sat. PD}  & Zero-filling & 32.34$\pm$2.83     & 0.80$\pm$0.11    & 30.47$\pm$2.71     & 0.74$\pm$0.14     \\ \cline{2-6} 
                                      & $\ell1$-SPIRiT    & 34.57$\pm$3.32 &0.81$\pm$0.11   & 31.51$\pm$2.21     & \textbf{0.78$\pm$0.08}                   \\ \cline{2-6} 
                                      & VN           & 35.83$\pm$4.43     & \textbf{0.84$\pm$0.13}    & 32.90$\pm$4.66     & \textbf{0.78$\pm$0.15}     \\ \cline{2-6} 
                                      & VS-Net       & \textbf{36.00$\pm$3.83}     & \textbf{0.84$\pm$0.13}    & \textbf{33.24$\pm$3.44}     & \textbf{0.78$\pm$0.15}     \\ \hline \hline

\multirow{4}{*}{Coronal PD}           & Zero-filling & 31.35$\pm$3.84     & 0.87$\pm$0.11    & 29.39$\pm$3.81     & 0.84$\pm$0.13     \\ \cline{2-6} 
                                      & $\ell1$-SPIRiT & 39.38$\pm$2.16 &0.93$\pm$0.03      & 34.06$\pm$2.41     & 0.88$\pm$0.04                   \\ \cline{2-6} 
                                      & VN           & 40.14$\pm$4.97     & 0.94$\pm$0.12    & 36.01$\pm$4.63     & 0.90$\pm$0.13     \\ \cline{2-6} 
                                      & VS-Net       & \textbf{41.27$\pm$5.25}     & \textbf{0.95$\pm$0.12}    & \textbf{36.77$\pm$4.84}     & \textbf{0.92$\pm$0.14}     \\ \hline \hline

\multirow{4}{*}{Axial fat-sat. T$_2$}    & Zero-filling & 36.47$\pm$2.34     & 0.94$\pm$0.02    & 34.90$\pm$2.39     & 0.92$\pm$0.02     \\ \cline{2-6} 
                                      & $\ell1$-SPIRiT & 39.38$\pm$2.70 &0.94$\pm$0.03      & 35.44$\pm$2.87     & 0.91$\pm$0.03               \\ \cline{2-6} 
                                      & VN           & 42.10$\pm$1.97     & \textbf{0.97$\pm$0.01}    & 37.94$\pm$2.29     & \textbf{0.94$\pm$0.02}     \\ \cline{2-6} 
                                      & VS-Net       & \textbf{42.34$\pm$2.06}     & 0.96$\pm$0.01    & \textbf{39.40$\pm$2.10}     & \textbf{0.94$\pm$0.02}     \\ \hline \hline

\multirow{4}{*}{Sagittal fat-sat. T$_2$} & Zero-filling & 37.35$\pm$2.69     & 0.93$\pm$0.07    & 35.25$\pm$2.68     & 0.90$\pm$0.09     \\ \cline{2-6} 
                                      & $\ell1$-SPIRiT & 41.27$\pm$2.95 &0.94$\pm$0.06      & 36.00$\pm$2.67     & 0.92$\pm$0.05             \\ \cline{2-6} 
                                      & VN           & 42.84$\pm$3.47     & \textbf{0.95$\pm$0.07}    & 38.92$\pm$3.23    & \textbf{0.93$\pm$0.09}     \\ \cline{2-6} 
                                      & VS-Net       & \textbf{43.10$\pm$3.44}     & \textbf{0.95$\pm$0.07}    & \textbf{39.07$\pm$3.33}     & 0.92$\pm$0.09     \\ \hline \hline

\multirow{4}{*}{Sagittal PD}          & Zero-filling & 37.12$\pm$2.58     & 0.96$\pm$0.04    & 35.96$\pm$2.57     & 0.94$\pm$0.05     \\ \cline{2-6} 
                                      & $\ell1$-SPIRiT & 44.52$\pm$1.94 &0.97$\pm$0.02      & 39.14$\pm$2.12     & 0.96$\pm$0.02               \\ \cline{2-6} 
                                      & VN           & 46.34$\pm$2.75     & \textbf{0.98$\pm$0.05}    & 39.71$\pm$2.58     & \textbf{0.96$\pm$0.05}     \\ \cline{2-6} 
                                      & VS-Net       & \textbf{47.22$\pm$2.89}     & \textbf{0.98$\pm$0.04}    & \textbf{40.11$\pm$2.46}     & \textbf{0.96$\pm$0.05}     \\ \hline
\end{tabular}}
\label{tb:compare}
\vspace{-20pt}
\end{table}

%\input{sections/results}
% !TeX root=../paper.tex

\section{Conclusion}
\label{sec:conclusion}
In this paper, we proposed the variable spitting network (VS-Net) for accelerated reconstruction of parallel MR images. We have detailed how to formulate VS-Net as an iterative energy minimization process embedded in a deep learning framework, where each stage essentially corresponds to one iteration of an iterative reconstruction. In experiments, we have shown that the performance of VS-Net gradually plateaued as the network stage number increased, and that setting parameters in each stage as learnable improved the quantitative results. Further, we have evaluated VS-Net on a multi-coil knee image dataset for 4-fold and 6-fold acceleration factors under Cartesian undersampling and showed its superiority over two state-of-the-art methods.

\section{Acknowledgements}
This work was supported by the EPSRC Programme Grant (EP/P001009/1) and the British Heart Foundation (NH/17/1/32725). The TITAN Xp GPU used for this research was kindly donated by the NVIDIA Corporation.

%\subsubsection*{Acknowledgements.} 
%The research was supported by the British Heart Foundation (NH/17/1/32725); National Institute for Health Research (NIHR) Biomedical Research Centre based at Imperial College Healthcare NHS Trust and Imperial College London; and the Medical Research Council, UK. We would like to thank Dr Simon Gibbs, Dr Luke Howard and Prof Martin Wilkins for providing the CMR image data. The Titan Xp for this research was donated by NVIDIA. 

\bibliographystyle{splncs}
\bibliography{refs}

\end{document}